\documentclass[10pt,aps,twocolumn,superscriptaddress,nofootinbib,tightenlines,floatfix,longbibliography]{revtex4-1}
\usepackage{amsmath,amssymb}
\usepackage{graphicx}
\usepackage{color}
\usepackage{comment}
\usepackage{bbold}
\usepackage{soul,color}
\sethlcolor{white}
\usepackage{booktabs}
\usepackage{array}
\usepackage{hyperref}
\hypersetup{
    colorlinks=true,       
    linkcolor=blue,          
    citecolor=blue,        
    filecolor=blue,      
    urlcolor=blue           
}

\def\Al{$^{26}$Al}
\def\Cl{$^{34}$Cl}

\begin{document}

\title{Effective Stellar $\beta$-Decay Rates of Nuclei with Long-lived Isomers: $^{26}$Al and $^{34}$Cl}

\author{Projjwal Banerjee}
\email{projjwal@sjtu.edu.cn}
\affiliation{Department of Astronomy, School of Physics and Astronomy, Shanghai Jiao Tong University, Shanghai 200240, China}

\author{G. Wendell Misch}
\email[Preferred: ]{wendell.misch@gmail.com}
\email{wendell@sjtu.edu.cn}
\affiliation{School of Physics and Astronomy, Shanghai Jiao Tong University, Shanghai 200240, China}
\affiliation{Collaborative Innovation Center of IFSA, Shanghai Jiao Tong University, Shanghai 200240, China}

\author{Surja K. Ghorui}
\email{surja@sjtu.edu.cn}
\affiliation{School of Physics and Astronomy, Shanghai Jiao Tong University, Shanghai 200240, China}

\author{Yang Sun}
\email{sunyang@sjtu.edu.cn}
\affiliation{School of Physics and Astronomy, Shanghai Jiao Tong University, Shanghai 200240, China}
\affiliation{Collaborative Innovation Center of IFSA, Shanghai Jiao Tong University, Shanghai 200240, China}
\affiliation{Institute of Modern Physics, Chinese Academy of Sciences, Lanzhou 730000, China}

\begin{abstract}
Isotopes with low-lying long-lived isomers can behave very differently from other isotopes in astrophysical environments. In particular, the assumption of thermal equilibrium in computing the temperature-dependent $\beta$-decay rates of such isotopes can fail below certain temperatures. We focus on the $\beta$-decay of $^{26}$Al since it is one of the most important isotopes in observational astrophysics and has a low-lying isomeric state; we compare and contrast these results with $^{34}$Cl.  We rule out recently reported $^{26}$Al effective $\beta$-decay rates that showed large differences from previous calculations, finding that we agree with the earlier results.  We conclude that in general, effective $\beta$-decay rates should be defined separately for the ground and isomeric states at temperatures where thermal equilibrium cannot be achieved.

\end{abstract}

\maketitle

\section{\label{intro}Introduction}

Nuclear structure effects are crucial inputs for calculating nuclear reaction rates in astrophysical conditions. These rates are used in nuclear reaction network codes to calculate the nucleosynthesis that occurs in hot stellar environments.  To this end, thermally averaged reaction rates are commonly used where each isotope that participates in nucleosynthesis is treated as a single species with the implicit assumption of a thermal equilibrium population of all excited nuclear states.

Isomers, however, can pose a major challenge to the thermal equilibrium assumption. An isomer is an excited nuclear state in which nuclear structure effects inhibit $\gamma$-decay to lower-lying states, endowing the isomer with a lifetime much longer than most nuclear states. The low transition rate from an isomeric state to lower-lying states is due to either transition selection rules for allowed transitions (as in spin isomers that require large changes of angular momentum and K isomers that must change their spin orientation relative to the nucleus's axis of symmetry) or energy barriers from nuclear structure effects (as in shape isomers that must change shape) \cite{wd:1999}. Because of the resultant weak coupling to the ground state, thermal equilibrium in stellar conditions can be unrealizable for isotopes with low-lying isomers, particularly if the isomeric state has a $\beta$-decay rate vastly different from the ground state. In these situations, it becomes particularly tricky to accurately treat the nuclear species in reaction network calculations.

The best-known example of this is the $\beta$-decay of {\Al}, which has a half-life against $\beta$-decay of $0.717$ Myr in the ground state (GS) and a long-lived isomeric state (IS) at 228 keV with a $\beta$-decay half-life of 6.35 s \cite{as:2005}. In astrophysics, {\Al} is an observationally important isotope. {\Al} was shown to be present in the early solar system via meteoritic excess of $^{26}$Mg \cite{lee1976}. {\Al} has since become a key isotope for the study of the formation and evolution of the early solar system using meteorites.  Additionally, 1.809 MeV gamma rays from decay of the first excited state of $^{26}$Mg produced by {\Al} $\beta$-decay is found extensively throughout the galaxy and provides critical information about ongoing star formation \cite{mahoney1982,diehl1995}.

Because of its observational importance, it is crucial to accurately calculate the synthesis of {\Al} in stars. Therefore, the effective $\beta$-decay (EBD) rates of {\Al} in stellar environments have been studied extensively \cite{wf:1980,coc2000,runkle2001,gupta2001,iliadis2011}.  Recent calculations using a novel formalism by Ref. \cite{reifarth} found EBD rates for {\Al} that deviate dramatically from the currently accepted rates \cite{coc2000,runkle2001} at temperatures greater than 40 keV.  This potentially has major implications for the yield of {\Al} produced in stars.

In this paper, we use a simple but precise method to compute the EBD rates of {\Al} in stellar conditions. We find that our rates agree very well with previous results by Ref. \citep{coc2000}, and we do not find the deviation reported by Ref.~\cite{reifarth}. We also compute the EBD rates for {\Cl} and find that the currently accepted rates are accurate. We clarify the definition of EBD rates for isotopes with low-lying isomers and show that they can be used for low and high temperatures. We point out limitations of EBD rates -- particularly at intermediate temperatures -- and discuss methods to treat them more accurately in nucleosynthesis calculations.

\section{Effective $\beta$-decay rate}

EBD rates are used extensively in stellar nucleosynthesis network codes. In stellar conditions, excited states become thermally populated. Since the excited states generally have $\beta$-decay rates different from the ground state, the effective rate of $\beta$-decay for the isotope differs from the ground state rate. The effective rate depends on temperature $T$, as the thermal occupation probability of exited states changes with $T$. Ideally, every excited state with an appreciable thermal population at a given temperature should be treated separately, i.e. essentially as a separate species in the nuclear reaction network. However, such an approach can greatly increase the size of the network, making the calculations computationally expensive.  The goal of using EBD rates is to reduce the network size by employing a single effective rate at a given temperature and density that accounts for the contribution from the ground state as well as the excited states.  This allows an isotope to be treated as a single species in the reaction network.

The most common approach to calculating $\beta$-decay is to assume that the nuclear states are in a thermal equilibrium distribution \cite{ffn:1980,ty1987,oda-etal:1994,lm:2001,msf:2018}. The thermal EBD rate is
\begin{equation}
\lambda_{eff}^{\beta}(T)=\sum\limits_{i} n_i(T)\lambda_i^\beta,
\label{eq:beff}
\end{equation}
where $n_i$ and $\lambda_i^\beta$ are the thermal occupation probability and $\beta$-decay rate of state $i$, respectively. Usually, $n_i$ increases with temperature for the excited states, and their contribution to $\lambda_{eff}^{\beta}$ increases accordingly. The values of $n_i$ come from the Boltzmann distribution.
\begin{equation}
    n_i(T) = \frac{2J_i+1}{G(T)}e^{-E_i/T}
\end{equation}
Here, $J_i$ and $E_i$ are the spin and energy of state $i$, respectively, and  $G(T)$ is the nuclear partition function at temperature $T$.

This method of calculating $\lambda_{eff}^{\beta}$ works well for most isotopes and is broadly used. For example, $s$-process (slow neutron capture process, where all nuclei are near $\beta$-stability) nuclear reaction networks mostly use a thermal equilibrium $\lambda_{eff}^{\beta}$ \cite{ty1987}. The criterion which determines thermal equilibrium to be a valid assumption is that the internal transition (IT) rates which push each state to its thermal equilibrium population are much faster than the individual $\beta$-decay rates.  At temperatures much lower than the lowest excited state and in the absence of production, $n_i\approx 0$ for all excited states, so $\lambda_{eff}^{\beta}$ corresponds to the laboratory $\beta$-decay rate of the ground state. At high temperatures, the photon bath efficiently drives internal transitions that bring the nucleus to thermal equilibrium.

Although thermal equilibrium is usually a valid assumption, it can fail at some temperatures in isotopes with low-lying long-lived isomers.  The GS and IS may behave very differently, and the IT rates that keep them in equilibrium are inefficient at lower temperatures.  If the long-lived states (GS and IS) have very different $\beta$-decay rates and one of them is comparable to or faster than the connecting IT rates, defining a single EBD rate becomes ambiguous. Nevertheless, EBD rates can still be defined for the GS and IS individually, allowing the individual states to be evolved separately as described in the next section.

\section{Methods}

The time evolution of the nuclear state abundances $N_i$ (the number of nuclei of a given species in state $i$) is described by the coupled differential equations \cite{wf:1980}
\begin{equation}
    \dot{N_i} = \sum\limits_{j}\left(\lambda_{ji}N_j - \lambda_{ij}N_i\right) - \left(\sum\limits_d \lambda_{i}^d\right) N_i + \sum\limits_p P_{i}^p
    \label{eq:Ni_dot}
\end{equation}
where $\lambda_{ij}$ is the IT rate from state $i$ to state $j$, $\lambda_{i}^d$ is the destruction rate of state $i$ through external channel $d$, and $P_{i}^p$ is the total production rate of state $i$ through external channel $p$. Here we take all $P_{i}^p=0$ and consider only destruction via $\beta$-decay.

Starting from an initial abundance of states $N_i(t=0)$, EBD rates can be defined by time-evolving each $N_i$ according to Eq.~\ref{eq:Ni_dot} to find the time $\tau_{eff}$ such that the total isotopic abundance drops by a factor of $e$.
\begin{equation}
    \sum\limits_{i} N_i(\tau_{eff})=1/e\sum\limits_{i} N_i(t=0)
\label{eq:criteria}    
\end{equation}
Then at a given temperature $T$, we have $\lambda_{\beta}^{eff} = 1/\tau_{eff}$.

At high temperatures where thermal equilibrium is achieved, the IT rates are fast enough that $\tau_{eff}$ is independent of the initial distribution of the $N_i$. However, at intermediate and low temperatures, $\tau_{eff}$ depends on the initial conditions, and the EBD rates are sensible only for individual long-lived states (i.e., the GS and isomers).  Then for each long-lived state, $\lambda_{eff}^{\beta}$ can be calculated with the criterion in Eq. \ref{eq:criteria} by starting from a initial population where all of the nuclei are in that state. We use this method to compute $\lambda_{eff}^{\beta}$ separately for the GS and the IS.

This prescription gives the correct EBD rates for long-lived states with both fast and slow $\beta$-decay rates at low and high temperatures. However, at intermediate temperatures, it fails to estimate the EBD rates for the IS of both {\Al} and {\Cl}. This failure is due to the fact that the $\beta$-decay rates become time-dependent at intermediate temperatures. We discuss this is detail in Sec.~\ref{sec:disc}. We use the Eq. \ref{eq:criteria} prescription described above for calculating the EBD rates and compare our results with calculations from previous studies.

\section{Nuclear Physics Inputs \label{sec:nuclear}}

We used measured values for IT rates and $\beta$-decay rates taken from Ref. \cite{ENSDF} whenever available. However, for many of the possible transitions, experimental values are not available. We estimated unknown IT and $\beta$-decay rates with shell model calculations using the  USDB interaction \cite{Brown2006}. Since the USDB interaction was obtained by fitting both single particle energies and two-body matrix elements to experimental data in $sd$-shell nuclei, we expect accurate predictions of unknown transition strengths for nuclei in this mass region.

The Hamiltonian was diagonalized using the shell model codes NUSHELLX@MSU \cite{Brown2014} and OXBASH \cite{oxbash}.  We consulted Ref. \cite{Richter2008} for shell model parameter values.  We used the recommended effective charges, i.e., $e_{p}$= 1.36e and $e_{n}$=0.45e.  The Gamow-Teller $\beta$-decay strengths are quenched by a factor of 0.584. For M1 transitions, the free-nucleon $g$ factors are quenched by a  factor of 0.9. For higher multipole magnetic transitions (M3 and M5), we used bare values of $g$ factors. This allowed us to compute all spontaneous IT rates $\lambda_{ij}^s$ where $E_i> E_j$.  Including stimulated emission, the transition rates $\lambda_{ij}$ for $E_i> E_j$ are \cite{runkle2001}
\begin{equation}
    \lambda_{ij}= \frac{\lambda_{ij}^s}{1-e^{-(E_i-E_f)/T}}.
\end{equation}
The IT rates for transitions with $E_i< E_j$ were computed using detailed balance and are given by
\begin{equation}
     \lambda_{ij} = \frac{g_j}{g_i}\frac{\lambda_{ji}^s}{e^{(E_j-E_i)/T}-1}.
\end{equation}

\section{Results \label{sec:results}}

Using the IT and $\beta$-decay rates obtained as described above, we evolved $N_i$ according to equation Eq.~\ref{eq:Ni_dot} to calculate the EBD rates for the GS and IS by finding $\tau_{eff}$ as defined in Eq.~\ref{eq:criteria}. In order to calculate the EBD rate for each state $i$, the initial abundance of all states except state $i$ were set to zero.

\subsection{\Al}

\begin{figure}
\centerline{\includegraphics[width=\columnwidth]{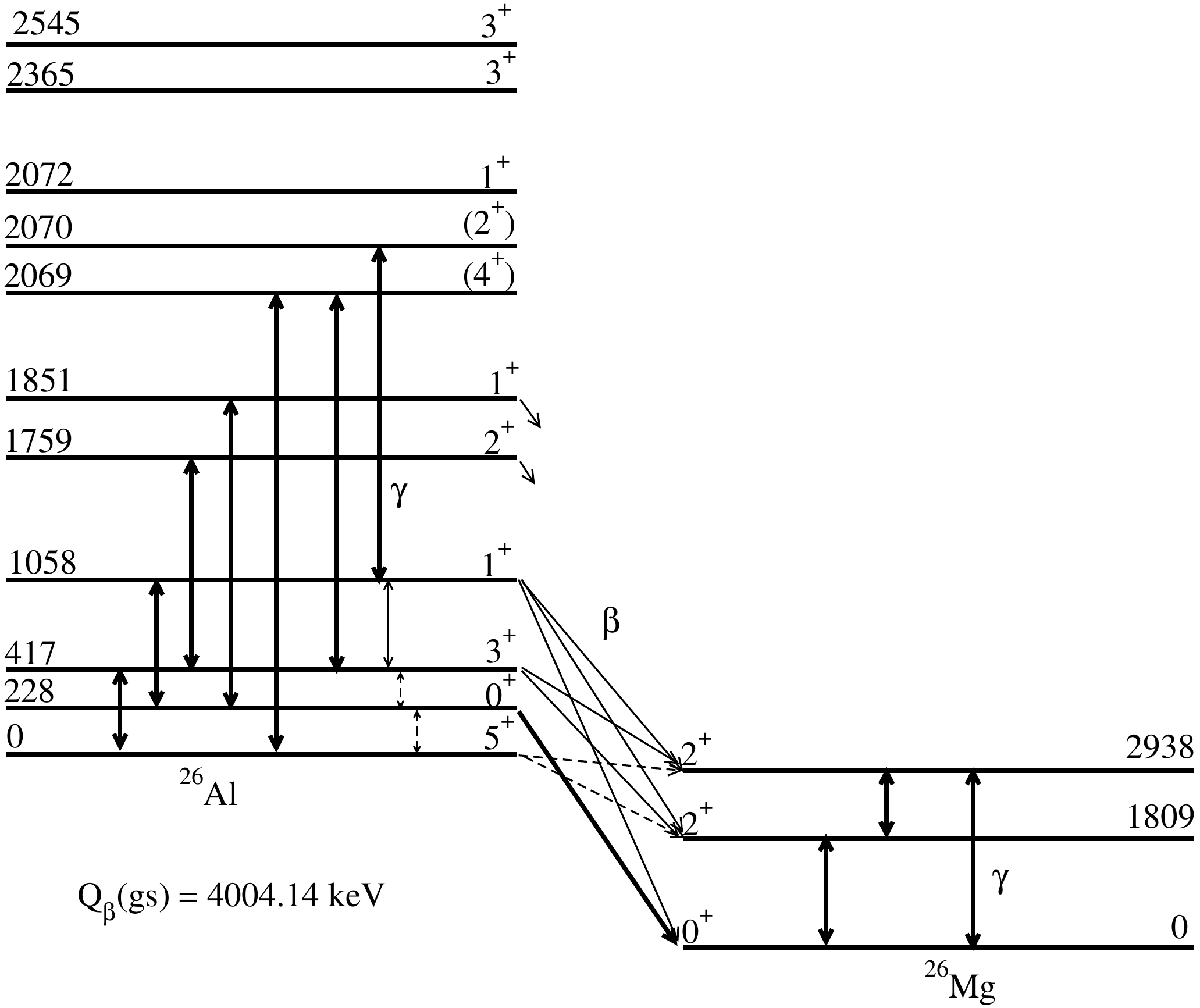}}
\caption{{\Al} level and decay diagram.  The state at 228 keV is a long-lived isomer.  Only low-lying $\gamma$ and $\beta$ transitions are shown.  The stronger and weaker transitions are represented by thick and thin arrows, respectively, and the broken arrows correspond to hindered transitions.  Experimental data are from Ref. \cite{ENSDF}.}
\label{fig:al26level}
\end{figure}

The {\Al} GS has a laboratory $\beta$-decay half-life of 0.717 Myr, while the IS has an energy of 228 keV and a $\beta$-decay half-life of 6.35 s.  The IS is directly connected only weakly to the GS via the highly-suppressed M5 transition shown in Fig. \ref{fig:al26level}.  This transition is still not observed in experiment; we estimate an M5 rate of $\lambda = 2.73\times 10^{-13}$ s$^{-1}$ from the shell model.  On the other hand, the GS couples strongly to the $3^{+}$ state at 417 keV (via an E2 transition) while the IS couples strongly to the $1^{+}$ state at 1058 keV (via an M1 transition).  An E2 transition connects the $3^{+}$ and $1^{+}$ states, and a weak M3 transition links the $3^+$ state to the IS (see Appendix \ref{apn:rates} for all transition rates).  Although weak, the ($1^+;1058$ keV $\rightarrow 3^+;417$ keV) E2 transition is the critical bridge between the GS and the IS at $T\lesssim 40$ keV. At very low temperatures the GS and the IS are essentially decoupled by thermal suppression of transitions from low to high energy, but the indirect coupling between them increases with increasing temperature via the $1^{+}$ and $3^{+}$ states.

\begin{figure}
\centerline{\includegraphics[width=\columnwidth]{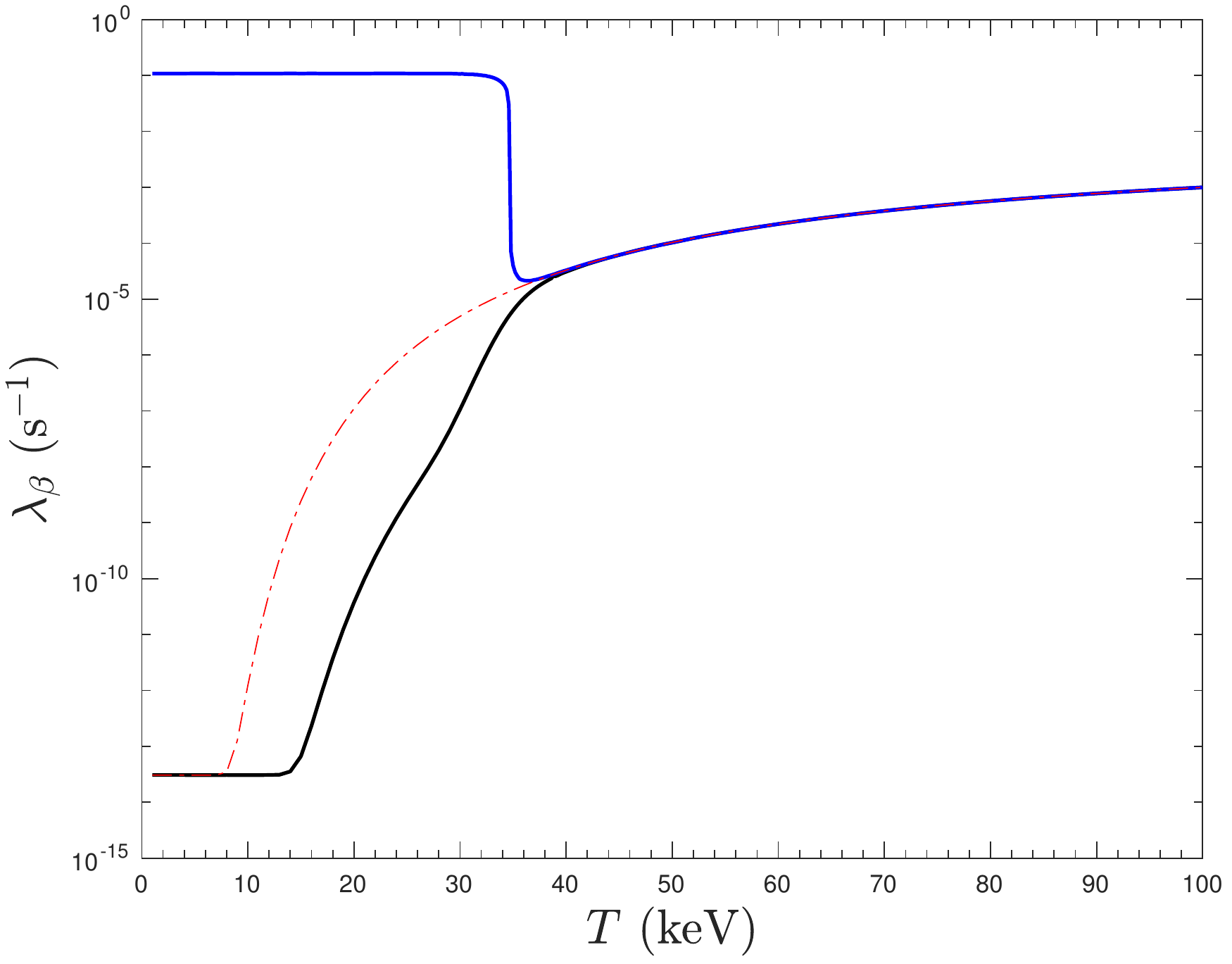}}
\caption{{\Al} $\beta$-decay rates calculated in this work.  \emph{Solid black:} ground state effective rate.  \emph{Solid blue:} isomeric state effective rate.  \emph{Dashed-dot red}: thermal equilibrium rate.}
\label{fig:al26eff}
\end{figure}

Fig.~\ref{fig:al26eff} shows the EBD rates for the GS and IS of {\Al}. For $T\lesssim 15$ keV, $\lambda_{eff}^\beta$ for both the GS and the IS are their respective laboratory $\beta$-decay rates. As $T$ increases however, the GS $\lambda_{eff}^\beta$ begins to increase because the IS starts to become thermally populated. Interestingly, although the $3^{+}$ and $1^{+}$ state are connected by a weaker E2 transition, its rate of $\sim 5\times 10^8~{\rm s}^{-1}$ is still sufficiently fast that it can act as an efficient bridge between the GS and the IS. Because the laboratory $\beta$-decay rate of the IS is greater than the GS by more than thirteen orders of magnitude, the GS $\lambda_{eff}^\beta$ diverges dramatically from the laboratory $\beta$-decay rate even at temperatures where the population of the IS is quite low. In contrast, the IS $\lambda_{eff}^\beta$ stays roughly equal to its laboratory rate up to $T\sim 30$ keV. Above this temperature, the IT rates become sufficiently high for the GS and IS to be strongly coupled via the higher lying $3^{+}$ and $1^{+}$ states. For $T\gtrsim 40$ keV, the $\lambda_{eff}^\beta$ is identical for both the GS and the IS and is equal to the thermal equilibrium rate. Thus, at temperatures where thermal equilibrium is quickly achieved, $\lambda_{eff}^\beta$ becomes well-defined for an isotope as a whole, since the rate is independent of the initial configuration. For $T\lesssim 40$ keV, $\lambda_{eff}^\beta$ applies only to individual states.

\begin{figure}
\centerline{\includegraphics[width=\columnwidth]{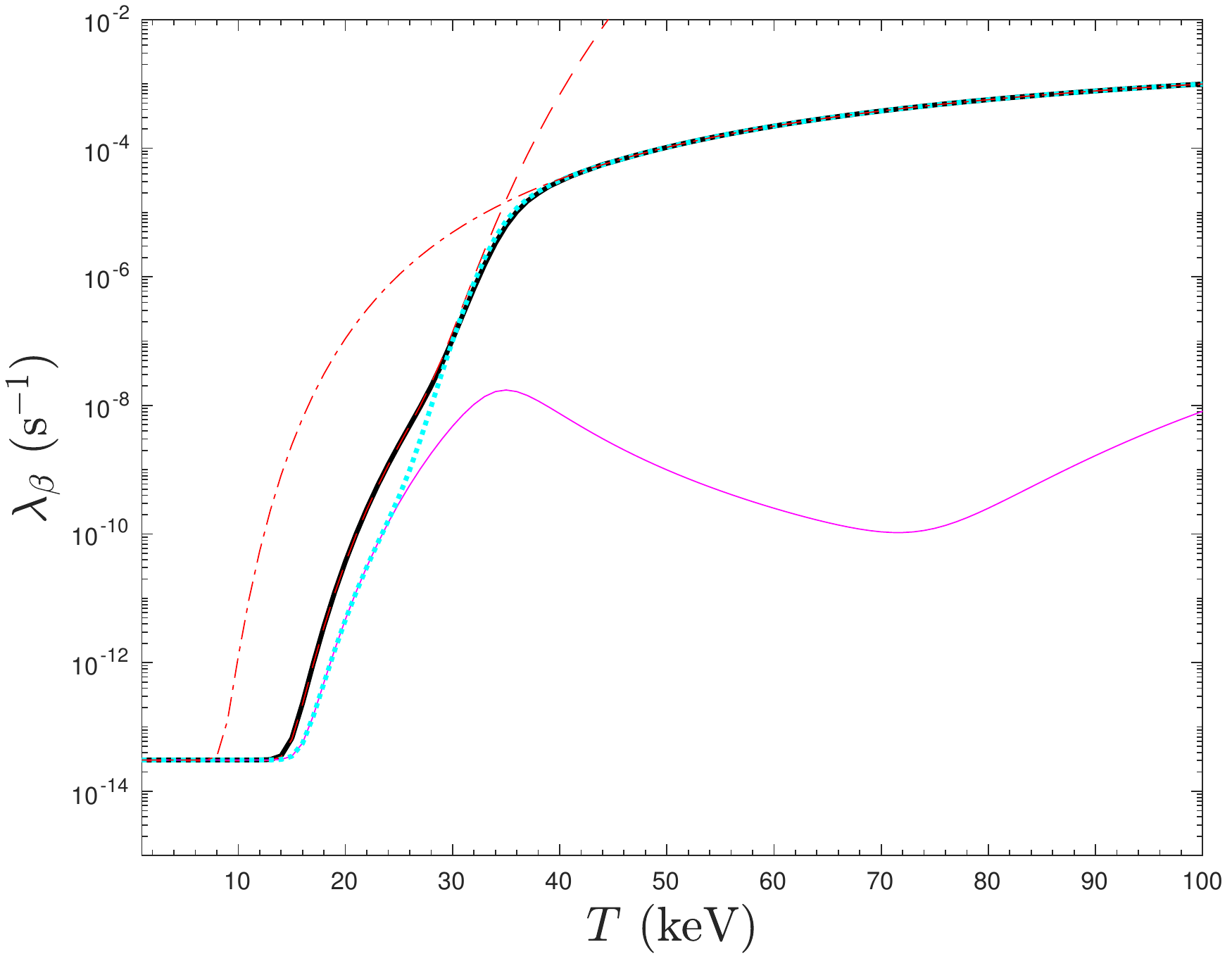}}
\caption{{\Al} $\beta$-decay rates from several calculations.  \emph{Solid black:} ground state effective rate as computed in this work.  \emph{Dotted cyan:} ground state effective rate as calculated in this work with Weisskopf rates for unknown internal transitions.  \emph{Dashed red:} off-equilibrium rate fit from Ref. \cite{coc2000}.  \emph{Solid magenta:} rate calculated by Ref.~\cite{reifarth}.  \emph{Dashed-dot red:} rate calculated assuming thermal equilibrium.}

\label{fig:al26comp}
\end{figure}

We compared our calculations with the standard accepted rates from Ref. \cite{coc2000} and recent results from Ref.~\cite{reifarth}; the former study reports the GS EBD rates as defined in Eq. \ref{eq:criteria}, whereas the latter employs a novel formalism.  As shown in Fig. \ref{fig:al26comp}, the rates calculated by Ref.~\cite{reifarth} deviate by orders of magnitude from Ref. \cite{coc2000} above $T\gtrsim 30$ keV, while our GS EBD rates match the older Ref. \cite{coc2000} rates.  Although Ref. \cite{reifarth} concluded that the inclusion of many {\Al} levels was responsible for this effect, we include the same states and do not find the large deviations at $T\gtrsim 30$ keV.   To investigate this, we repeated our calculations, replacing our shell model IT rates with the Weisskopf approximation used in Ref.~\cite{reifarth}.  We found that this only slightly changes the EBD rates for $T\lesssim 30$ keV and does not affect the rates at $T\gtrsim 30$ keV.

\subsection{\Cl}

\begin{figure}
\centerline{\includegraphics[width=\columnwidth]{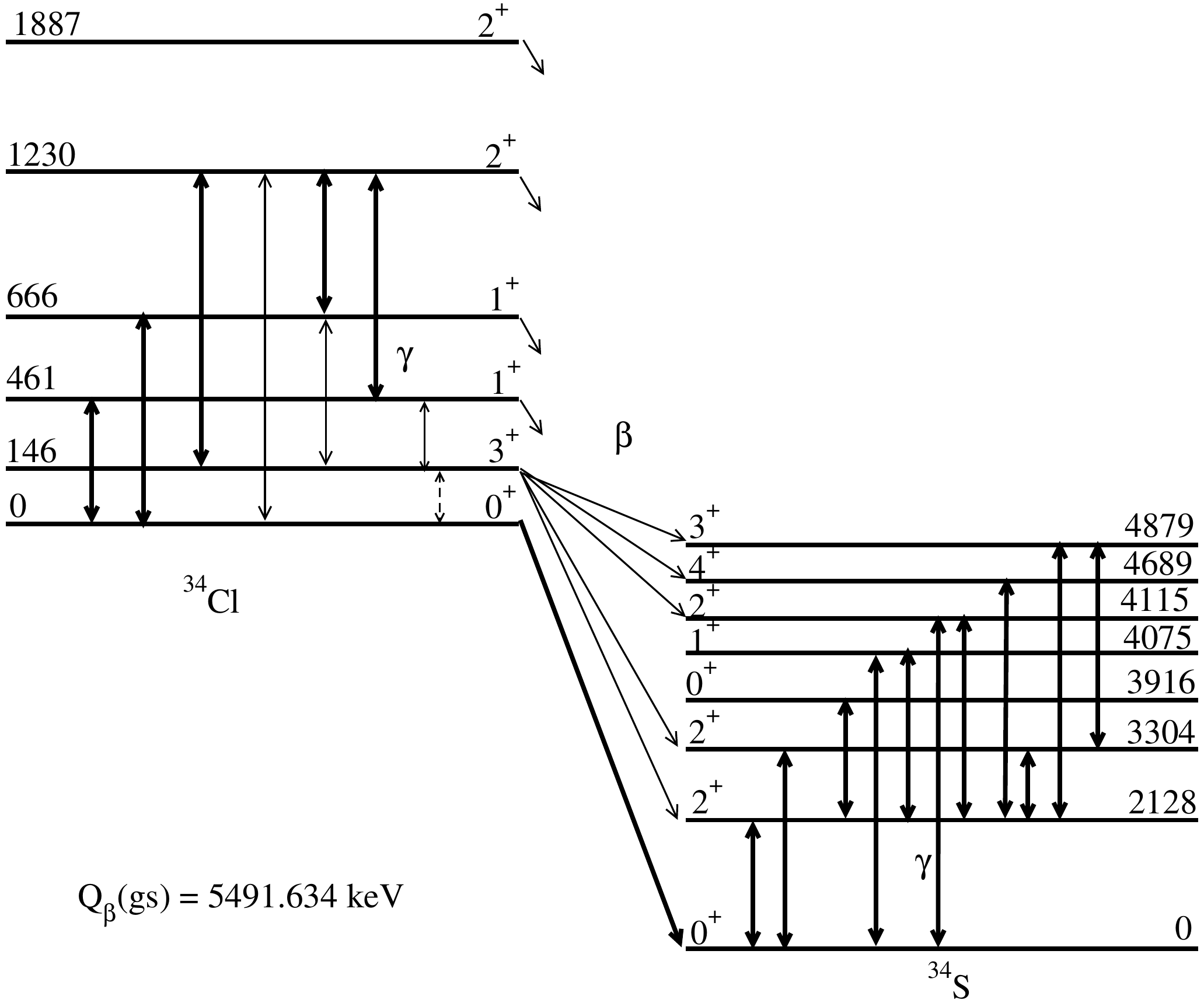}}
\caption{{\Cl} level diagram.  The state at 146 keV is a long-lived isomer.  The arrows are as in Fig. \ref{fig:al26level}.  Experimental data are from Ref. \cite{ENSDF}.}
\label{fig:cl34level}
\end{figure}

We computed the EBD rates for {\Cl} to further check our methods, comparing the results with previous studies. For this nucleus, we considered only the lowest six levels, shown in Fig. \ref{fig:cl34level}.

{\Cl} also has a low-lying isomer; its $\beta$-decay half-life is 58 min, long compared to the 1.53 s half-life of the GS. This contrasts with {\Al}, wherein the GS is much more $\beta$-stable than the IS. Furthermore, the {\Cl} IS is connected to the GS by an M3 transition, which is less forbidden than the M5 transition directly linking the {\Al} IS and GS. Nevertheless, the {\Cl} M3 transition constitutes only a weak direct coupling to the GS, so these states communicate more efficiently at intermediate temperatures via the $1^{+}$ state at 468 keV. This intermediate state strongly couples to the GS (M1 transition) and somewhat less strongly to the IS (E2 transition). These transition strengths are sufficient for the state to act as an efficient bridge at intermediate and high temperatures. The second $1^{+}$ state at 666 keV, being still fairly low in excitation energy, provides additional routes to couple the GS and the IS.

\begin{figure}
\centerline{\includegraphics[width=\columnwidth]{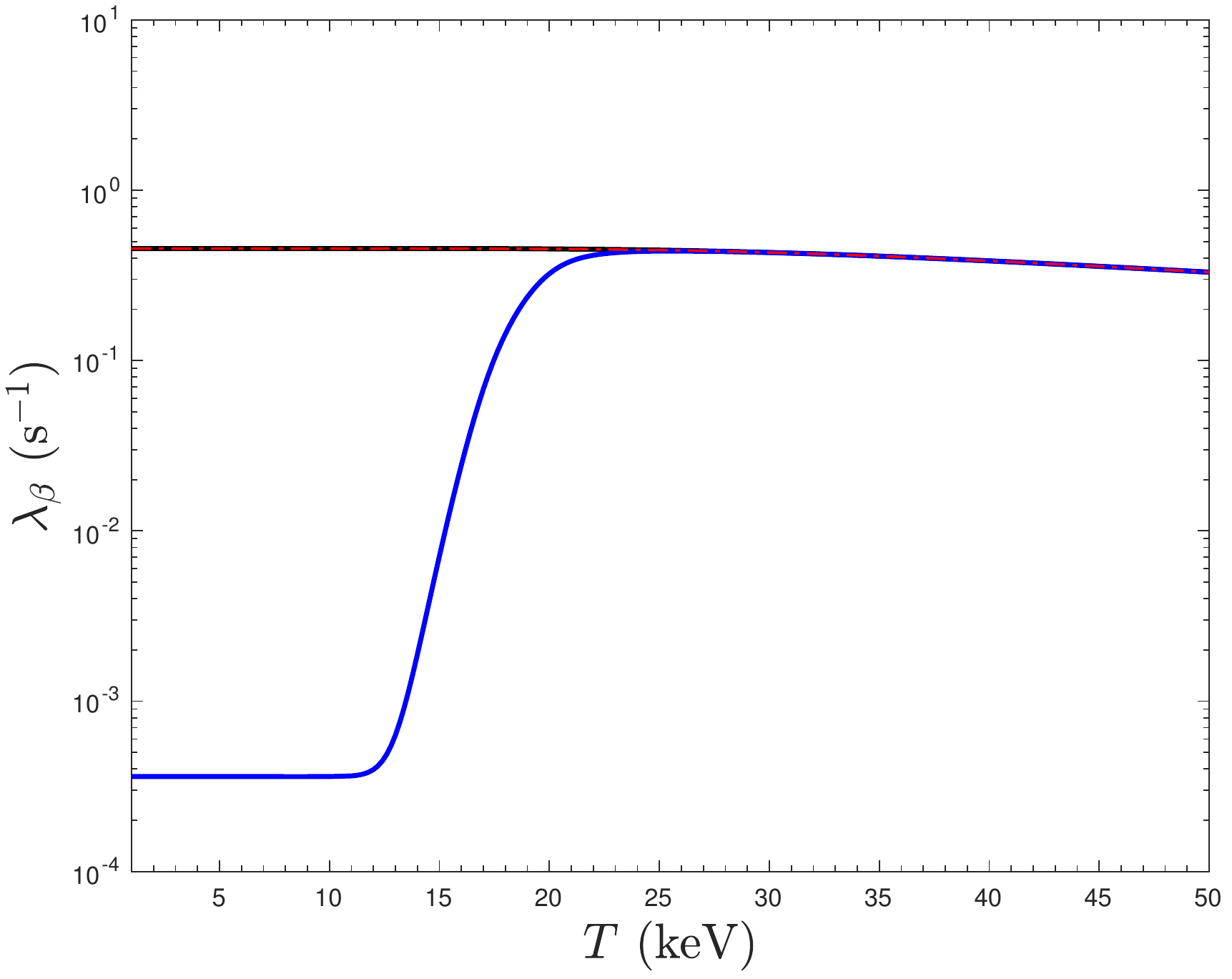}}
\caption{{\Cl} $\beta$-decay rates calculated in this work.  \emph{Solid black:} ground state effective rate.  \emph{Solid blue:} isomeric state effective rate.  \emph{Dashed-dot red}: thermal equilibrium rate.}
\label{fig:cl34eff}
\end{figure}

Figure \ref{fig:cl34eff} shows our EBD rates for the GS and the IS. As with {\Al} at high temperature, the EBD rates for both states are equal to the thermal equilibrium rate at $T\gtrsim 25$ keV. At low temperatures, the EBD rates correspond to the laboratory $\beta$-decay rates.  At all temperatures, the GS EBD rate is essentially equal to the thermal equilibrium rate, which in turn is very similar to the laboratory GS $\beta$-decay rate even at high temperature.  This is due to the fast laboratory GS $\beta$-decay rate which, when coupled with its high thermal population factor, makes the GS the dominant contributor to $\beta$-decay at all temperatures. This contrasts with the IS, where even a slight flow of population to the ground state has a large effect on the IS EBD rate.

\begin{figure}
\centerline{\includegraphics[width=\columnwidth]{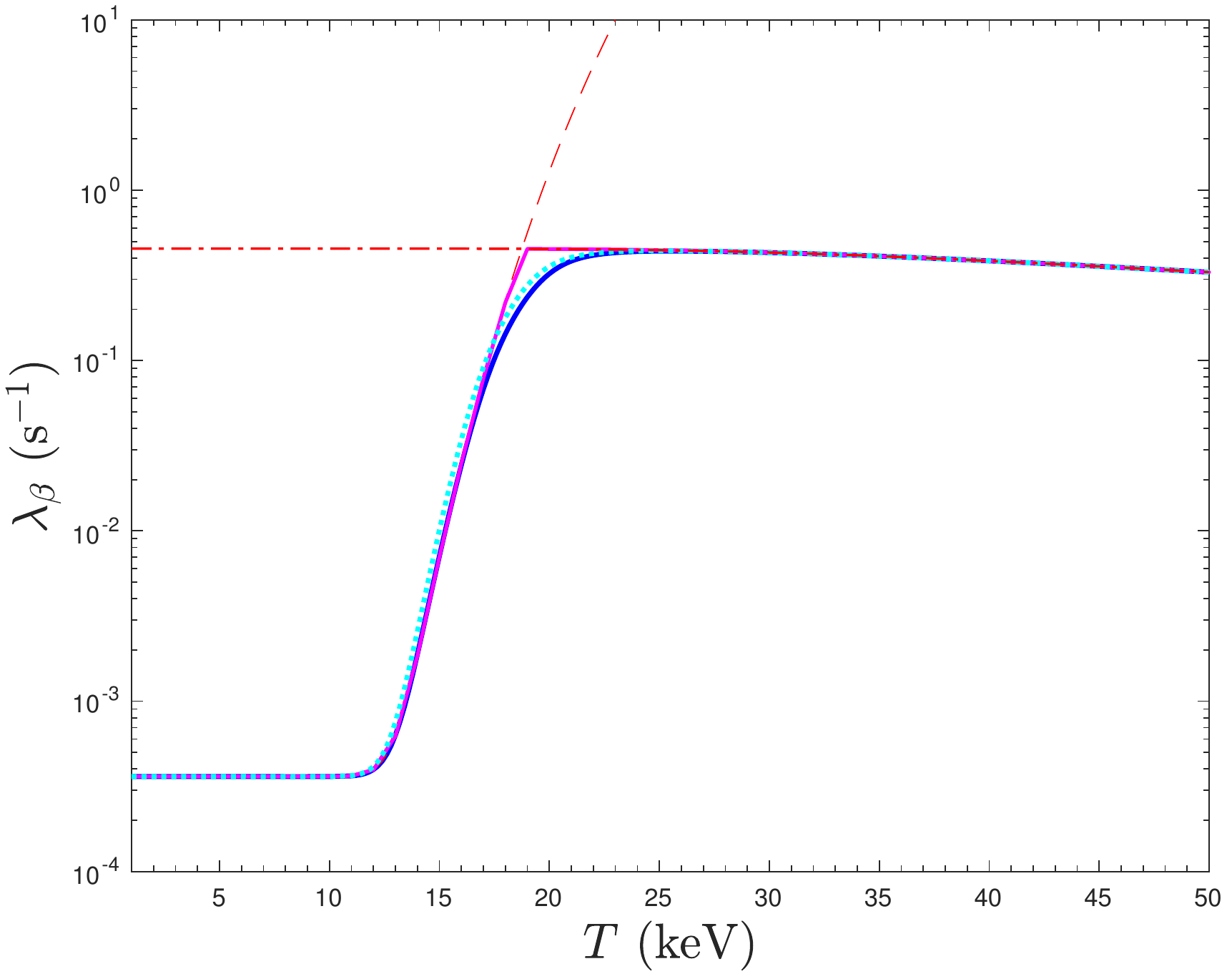}}
\caption{{\Cl} $\beta$-decay rates from several calculations.  \emph{Solid blue:} isomeric state effective rate as computed in this work.  \emph{Dotted cyan:} isomeric state effective rate as calculated in this work with Weisskopf rates for unknown internal transitions.  \emph{Dashed red:} off-equilibrium rate fit from Ref. \cite{coc2000}.  \emph{Solid magenta:} rate calculated by Ref.~\cite{reifarth}.  \emph{Dashed-dot red:} rate calculated assuming thermal equilibrium.}
\label{fig:cl34comp}
\end{figure}

Fig. \ref{fig:cl34comp} compares our rates with earlier results.  We find that our IS EBD rates match the rates from Refs. \cite{coc2000,reifarth} at high and low temperatures.

\section{Discussion \label{sec:disc}}

As we detailed above, the EBD rate for an isotope with a low-lying isomer is not well defined at temperatures below which the assumption of thermal equilibrium is valid. In such nuclei, the EBD rate can be computed from Eqs. \ref{eq:Ni_dot} and \ref{eq:criteria} for the individual long-lived states (the ground and isomeric states) for use in reaction network calculations.  These individual state EBD rates are valid at both low and high temperatures.  At low temperature, the IT rates are so slow that the GS and IS are essentially independent, and at high temperature, the IT rates rapidly bring all states to their thermal equilibrium populations.

\begin{figure}
\centerline{\includegraphics[width=\columnwidth]{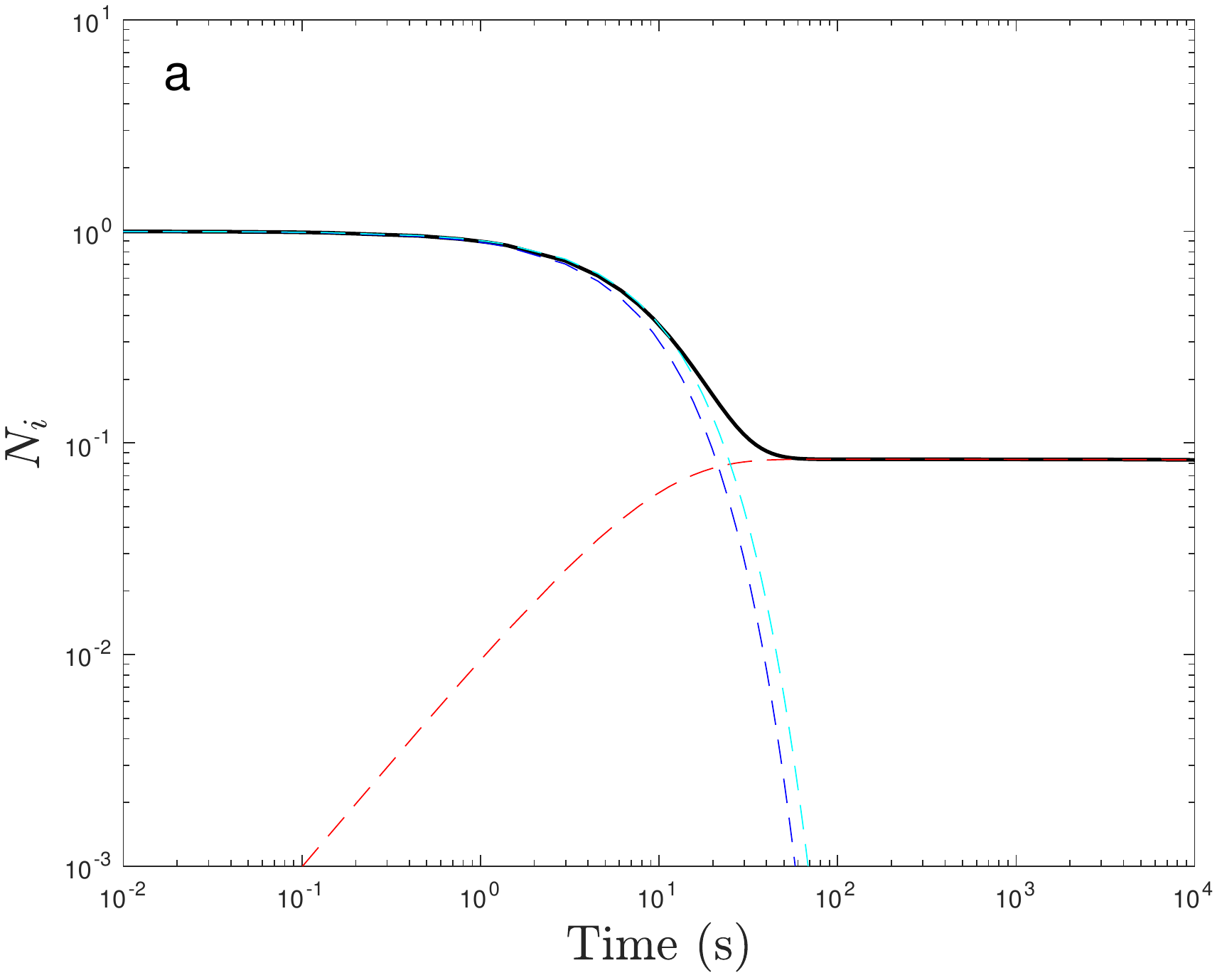}}
\centerline{\includegraphics[width=\columnwidth]{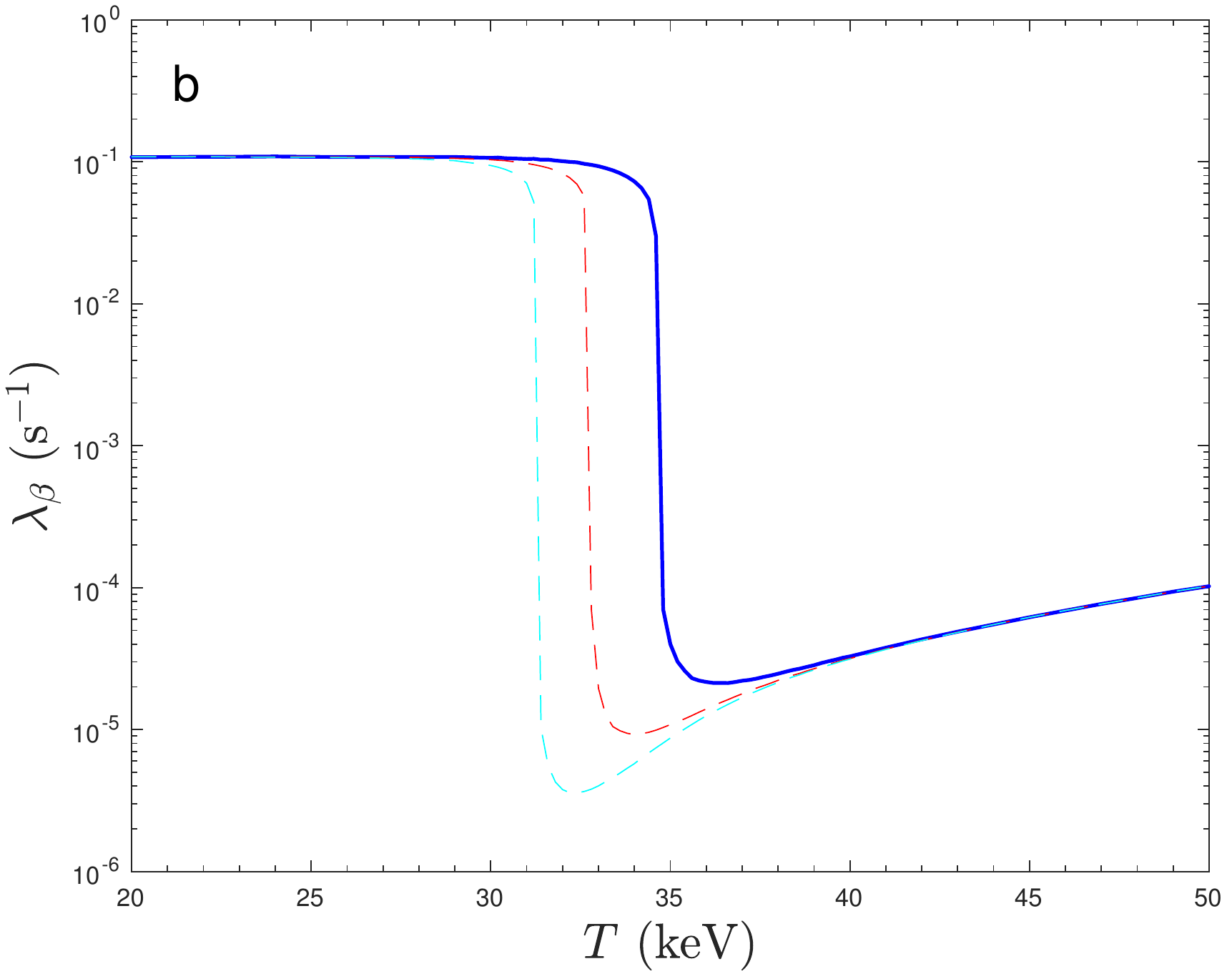}}
\caption{(a) Time evolution of {\Al} abundance at temperature $T=32$ keV starting with an initial population entirely in the isomeric state.  \emph{Solid black:} total abundance.  \emph{Dashed blue:} isomer abundance.  \emph{Dashed red:} ground state abundance.  \emph{Dashed cyan:} total abundance assuming the isomer effective $\beta$-decay rate.  (b) Effective $\beta$-decay rates estimated from the time required for the total abundance to decay by a factor of $e$ (\emph{solid blue}), $e^2$ (\emph{dashed red}), and $e^3$ (\emph{dashed cyan}).}
\label{fig:al26iso}
\end{figure}

At intermediate temperatures, this technique may give inaccurate results.  In {\Al}, for example, state population can flow from a fully populated IS to the more $\beta$-stable GS at a rate fast enough to impact the overall $\beta$-decay rate but too slow to drive the nucleus to thermal equilibrium.  The consequence is that the first $e$-folding in total abundance will take less time than the next.  While the EBD rate works well for the {\Al} GS, it thus fails for the IS at intermediate temperatures. This can be seen in Fig.~\ref{fig:al26iso}a where the total abundance (initially all in the IS) decays by a factor of $e$ in $\sim 10$ s before reaching a quasi-equilibrium steady state in $\sim 100$ s from which it decays slowly.  The transient behavior implies that the rate computed from Eq. \ref{eq:criteria} is too fast.  Fig. \ref{fig:al26iso}b emphasizes this fact by comparing {\Al} IS EBT rates estimated from the time required to decay by factors of $e, e^2,$ and $e^3$.  A similar problem arises for the IS of {\Cl}, although because the IS is \emph{more} $\beta$-stable than the GS, the transient behavior is a slower decay rate.  Plainly stated, the true $\beta$-decay rate in these examples evolves with time, so a single constant value loses its meaning.

The steady state attained at higher temperatures (thermal equilibrium) and at intermediate temperatures (quasi-equilibrium) is described by the population distribution $n_i$ that satisfies
\begin{equation}
    \dot{n_i}=0. 
\end{equation}
In fact, an EBD rate can also be defined using this steady state criterion. This quantity works well at higher temperatures when thermal equilibrium is attained, but at low temperatures it does not reflect the evolution of the nuclear species abundance in a star.  Stellar nuclear processes tend to preferentially produce the isotope in or near a particular long-lived state, so at intermediate temperatures in can run into problems similar to those discussed above for Eq. \ref{eq:criteria}.  Nevertheless, it is a useful concept for understanding how isotopes with isomers can behave in stellar conditions; we will explore this in a forthcoming paper \cite{steady}.

The difficulties with intermediate temperatures can pose a serious problem in stellar nucleosynthesis calculations if isotopes with isomers are produced at such temperatures.  The most straightforward way to accurately calculate the $\beta$-decay at intermediate temperatures is to treat the nuclear states of isotopes with isomers at separate species, but this of course would increase the network size and thus add to the computational cost.  On the other hand, given that post-processing nuclear reaction network calculations often include several hundred nuclides, adding a few more species for certain isotopes will not amount to a substantial inflation of computational costs.  In fact, although we included 11 states in our {\Al} calculations, Ref. \cite{iliadis2011} showed that using only the first four states will yield accurate results. Similarly, we found that the first three states are sufficient for {\Cl}.\\

\section{Conclusions \label{sec:concl}}

We calculated the effective $\beta$-decay (EBD) rates for {\Al} and {\Cl} at temperatures appropriate for stellar conditions.  Our results agree with previous calculations by Refs. \cite{coc2000,runkle2001,iliadis2011}, and we do not find the deviation in the high temperature EBD rates for {\Al} reported by Ref. \cite{reifarth}.  While we do not speculate on the reason for the discrepancy in Ref. \cite{reifarth}, we positively rule it out.

We point out that at temperatures below which thermal equilibrium is reachable for a particular nuclear species (typically species with long-lived isomers), the EBD rate is not well-defined for the isotope as a whole.  Instead, EBD rates should be defined separately for the ground state (GS) and the isomeric state (IS).  These rates can be used in nuclear reaction networks where the GS and IS are treated as two different species.  However, at intermediate temperatures, the EBD rate gives inaccurate results for the {\Al} IS and {\Cl} IS, although it works well for the GS of both nuclides.

It is important to note that particular isotopes are usually synthesized in stars at specific temperatures that may or may not be intermediate.  For example, {\Al} is synthesized either at low temperatures ($T\lesssim 10$ keV) during core H burning or at high temperatures ($T\gtrsim 90$ keV) during shell C/Ne burning \cite{limongi2006}.  The EBD rates will give accurate results at these temperatures provided the GS and the IS are treated separately.  Indeed, Ref. \cite{iliadis2011} showed the EBD rates give identical results when compared to calculations where the nuclear states are treated explicitly. However, since only a handful of isotopes with low-lying isomers are relevant for stellar nucleosynthesis and only a few states ($\lesssim 5$) are needed for the explicit treatment to produce accurate results, this generally more robust treatment for such isotopes may be worth the computational cost.

\section{Acknowledgments}

We thank George Fuller for helpful discussions.  Funding for this research came from the following: PB was supported by the National Natural Science Foundation of China (No. 11533006), and  GWM, SKG, and SY were supported by the National Natural Science Foundation of China (No. 11575112) and the National Key Program for S\&T Research and Development (No. 2016YFA0400501).

\bibliographystyle{h-physrev}
\bibliography{references}

\begin{thebibliography}{10}

\bibitem{wd:1999}
P.~Walker and G.~Dracoulis,
\newblock Nature {\bf 399}, 35 (1999).

\bibitem{as:2005}
A.~Aprahamian and Y.~Sun,
\newblock Nature Phys. {\bf 1}, 81 (2005).

\bibitem{lee1976}
T.~{Lee}, D.~A. {Papanastassiou}, and G.~J. {Wasserburg},
\newblock Geophys. Res. Lett. {\bf 3}, 109 (1976).

\bibitem{mahoney1982}
W.~A. {Mahoney}, J.~C. {Ling}, A.~S. {Jacobson}, and R.~E. {Lingenfelter},
\newblock \apj {\bf 262}, 742 (1982).

\bibitem{diehl1995}
R.~{Diehl} {\em et~al.},
\newblock Astron. Astrophys. {\bf 298}, 445 (1995).

\bibitem{wf:1980}
R.~A. {Ward} and W.~A. {Fowler},
\newblock \apj {\bf 238}, 266 (1980).

\bibitem{coc2000}
A.~{Coc}, M.-G. {Porquet}, and F.~{Nowacki},
\newblock \prc {\bf 61}, 015801 (2000).

\bibitem{runkle2001}
R.~C. {Runkle}, A.~E. {Champagne}, and J.~{Engel},
\newblock \apj {\bf 556}, 970 (2001).

\bibitem{gupta2001}
S.~S. {Gupta} and B.~S. {Meyer},
\newblock \prc {\bf 64}, 025805 (2001).

\bibitem{iliadis2011}
C.~{Iliadis}, A.~{Champagne}, A.~{Chieffi}, and M.~{Limongi},
\newblock Astrophys. J. Supp. {\bf 193}, 16 (2011), 1101.5553.

\bibitem{reifarth}
R.~Reifarth {\em et~al.},
\newblock Int. J. Mod. Phys. A {\bf 0}, 1843011 (0).

\bibitem{ffn:1980}
G.~M. {Fuller}, W.~A. {Fowler}, and M.~J. {Newman},
\newblock Astrophys. J. (Supplement) {\bf 42}, 447 (1980).

\bibitem{ty1987}
K.~{Takahashi} and K.~{Yokoi},
\newblock Atomic Data and Nuclear Data Tables {\bf 36}, 375 (1987).

\bibitem{oda-etal:1994}
T.~{Oda}, M.~{Hino}, K.~{Muto}, M.~{Takahara}, and K.~{Sato},
\newblock At. Data. Nucl. Data Tables {\bf 56}, 231 (1994).

\bibitem{lm:2001}
K.~{Langanke} and G.~{Mart{\'{\i}}nez-Pinedo},
\newblock At. Data. Nucl. Data Tables {\bf 79}, 1 (2001).

\bibitem{msf:2018}
G.~W. {Misch}, Y.~{Sun}, and G.~M. {Fuller},
\newblock \apj {\bf 852}, 43 (2018), 1708.08792.

\bibitem{ENSDF}
{Evaluated Nuclear Structure Data File database},
\newblock \url{http://www.nndc.bnl.gov/ensdf},
\newblock Accessed: 2018-03-28.

\bibitem{Brown2006}
B.~A. Brown and W.~A. Richter,
\newblock \prc {\bf 74} (2006).

\bibitem{Brown2014}
B.~Brown and W.~Rae,
\newblock Nuclear Data Sheets {\bf 120}, 115 (2014).

\bibitem{oxbash}
B.~A. {Brown} {\em et~al.},
\newblock MSU-NSCL Report No. 1289  (2004).

\bibitem{Richter2008}
W.~A. Richter, S.~Mkhize, and B.~A. Brown,
\newblock \prc {\bf 78}, 064302 (2008).

\bibitem{steady}
G.~W. {Misch}, P.~{Banerjee}, S.~K. {Ghorui}, and Y.~{Sun(to be published)}.

\bibitem{limongi2006}
M.~Limongi and A.~Chieffi,
\newblock \apj {\bf 647}, 483 (2006).

\end{thebibliography}

\appendix

\section{\label{apn:rates} Electromagnetic and $\beta$-decay rates}

We provide here our experimental and shell model IT and $\beta$-decay rate inputs.  We show experimental (exp) and shell model (SM) electromagnetic rates side-by-side for comparison.  Wherever available, we use the experimental values in our calculations.

\begin{table*}
\centering
\caption{Electromagnetic transition rates and $\beta$-decay rates for {\Al}.  Starred (*) $\beta$-decay rates are experimental values.  All experimental energy levels and transition/decay rates were taken from Ref. \cite{ENSDF}.}
\label{tab:al26emrate}
\begin{tabular}{lcr|lclcr|r}
\hline\hline
E$_{i}$ (keV)      & J$_{i}\pi$ & $\lambda_i^\beta$ (s$^{-1}$) & E$_{f}$ (keV)       & J$_{f}\pi$ & E$_\gamma$ (keV)& Multipolarity & $\lambda_{if}^{\rm exp}$ (s$^{-1}$)     & $\lambda_{if}^{\rm SM}$ (s$^{-1}$) \\
\hline
0.0      & 5+   & *3.07E-14 & -- & -- & -- & -- & -- & -- \\
228.305  & 0+   & *1.09E-01 & 0.0      & 5+   & 228.305  & M5 & --       & 2.73E-13 \\
416.852  & 3+   & 1.09E-04 & 0.0      & 5+   & 416.848  & E2 & 5.58E+08 & 8.16E+08 \\
         &      &  & 228.305  & 0+   & 188.547  & M3 & --       & 6.57E-02   \\
1057.739 & 1+   & 9.88E-02 & 0.0      & 5+   & 1057.739 & E4 & --       & 1.53E-02 \\
         &      &  & 228.305  & 0+   & 829.30   & M1 & 2.77E+13 & 3.17E+13 \\
         &      &  & 416.852  & 3+   & 640.887  & E2 & --       & 5.24E+08 \\
1759.034 & 2+   & 4.81E-03 & 0.0      & 5+   & 1759.034 & M3 & --       & 3.62E+04 \\
         &      &  & 228.305  & 0+   & 1530.729 & E2 & --       & 1.01E+08 \\
         &      &  & 416.852  & 3+   & 1342.145 & E2 & 1.79E+11 & 5.54E+10 \\
         &      &  & 1057.739 & 1+   & 701.285  & E2 & 3.58E+09 & 1.37E+09 \\
1850.62  & 1+   & 1.06E-01 & 0.0      & 5+   & 1850.62  & E4 & --       & 4.87E+01 \\
         &      &  & 228.305  & 0+   & 1622.0 7 & M1 & 2.15E+13 & 2.69E+13 \\
         &      &  & 416.852  & 3+   & 1433.73  & E2 & 1.51E+11 & 4.20E+12 \\
         &      &  & 1057.739 & 1+   & 792.881  & E2 & --       & 2.56E+08 \\
         &      &  & 1759.034 & 2+   & 91.586   & E2 & --       & 1.19E+05 \\
2068.86  & (4+) & 3.70E-05 & 0.0      & 5+   & 2068.77  & E2 & 6.93E+11 & 8.59E+11 \\
         &      &  & 228.305  & 0+   & 1840.555 & E4 & --       & 1.68E-02 \\
         &      &  & 416.852  & 3+   & 1651.95  & E2 & 1.54E+12 & 8.59E+11 \\
         &      &  & 1057.739 & 1+   & 1011.121 & M3 & --       & 9.39E+01 \\
         &      &  & 1759.034 & 2+   & 309.826  & E2 & --       & 5.63E+06 \\
         &      &  & 1850.62  & 1+   & 218.24   & M3 & --       & 1.12E-03 \\
2069.47  & (2+) & 1.13E-01 & 0.0        & 5+   & 2069.47  & M3 & --       & 1.01E+05 \\
         &      &  & 228.305  & 0+   & 1841.09  & E2 & 1.49E+12 & 1.49E+12 \\
         &      &  & 416.852  & 3+   & 1652.56  & M1 & 1.05E+13 & 1.66E+13 \\
         &      &  & 1057.739 & 1+   & 1011.71  & M1 & 3.73E+13 & 5.72E+13 \\
         &      &  & 1759.034 & 2+   & 310.43   & M1 & 8.58E+10 & 1.32E+11 \\
         &      &  & 1850.62  & 1+   & 218.85   & M1 & 1.98E+10 & 3.70E+10 \\
         &      &  & 2068.86  & (4+) & 0.61     & E2 & --       & 2.67E-09 \\
2071.64  & 1+   & 2.13E-02 & 0.0      & 5+   & 2071.64  & E4 & --       & 5.63E-01  \\
         &      &  & 228.305  & 0+   & 1842.8   & M1 & 5.51E+09 & 1.04E+13 \\
         &      &  & 416.852  & 3+   & 1654.73  & E2 & 6.55E+08 & 1.20E+09 \\
         &      &  & 1057.739 & 1+   & 1013.901 & E2 & --       & 2.17E+09 \\
         &      &  & 1759.034 & 2+   & 312.606  & E2 & --       & 1.45E+06 \\
         &      &  & 1850.62  & 1+   & 221.02   & E2 & 2.48E+06 & 1.19E+06 \\
         &      &  & 2068.86  & (4+) & 2.78     & M3 & --       & 1.83E-16 \\
         &      &  & 2069.47  & (2+) & 2.17     & E2 & --       & 2.78E-06 \\
2365.15  & 3+   & 1.73E-03 & 0.0      & 5+   & 2365.034 & E2 & 7.73E+09 & 1.69E+09 \\
         &      &  & 228.305  & 0+   & 2136.845 & M3 & --       & 2.94E+04 \\
         &      &  & 416.852  & 3+   & 1948.219 & E2 & 2.86E+11 & 1.19E+11 \\
         &      &  & 1057.739 & 1+   & 1307.375 & E2 & 1.19E+11 & 5.16E+10 \\
         &      &  & 1759.034 & 2+   & 606.108  & E2 & 1.28E+10 & 9.78E+09 \\
         &      &  & 1850.62  & 1+   & 514.53   & E2 & --       & 5.29E+08 \\
         &      &  & 2068.86  & (4+) & 296.29   & E2 & --       & 2.51E+07 \\
         &      &  & 2069.47  & (2+) & 295.678  & M1 & 4.42E+11 & 9.05E+09 \\
         &      &  & 2071.64  & 1+   & 293.51   & E2 & --       & 2.13E+06 \\
2545.367 & 3+   & 3.29E-03 & 0.0      & 5+   & 2545.232 & E2 & 2.12E+09 & 5.39E+09 \\
         &      &  & 228.305  & 0+   & 2317.062 & M3 & --       & 1.39E+06 \\
         &      &  & 416.852  & 3+   & 2128.421 & M1 & 2.61E+11 & 2.12E+11 \\
         &      &  & 1057.739 & 1+   & 1487.582 & E2 & 2.53E+10 & 7.59E+07 \\
         &      &  & 1759.034 & 2+   & 786.320  & E2 & 3.35E+10 & 7.94E+07 \\
         &      &  & 1850.62  & 1+   & 694.747  & E2 & --       & 5.31E+08 \\
         &      &  & 2068.86  & (4+) & 476.507  & E2 & --       & 4.40E+08 \\
         &      &  & 2069.47  & (2+) & 475.892  & M1 & 6.83E+11 & 2.28E+12 \\
         &      &  & 2071.64  & 1+   & 473.727  & E2 & --       & 1.32E+08 \\
         &      &  & 2365.15  & 3+   & 180.217  & E2 & --       & 2.58E+05\\
         \hline\hline
\end{tabular}
\end{table*}

\begin{table*}
\centering
\caption{Electromagnetic transition rates and $\beta$-decay rates for {\Cl}.  Starred (*) $\beta$-decay rates are experimental values.  All experimental energy levels and transition/decay rates were taken from Ref. \cite{ENSDF}.}
\label{tab:cl34emrate}
\begin{tabular}{lcr|lclcr|r}
\hline\hline
E$_{i}$ (keV)      & J$_{i}\pi$  & $\lambda_i^\beta$ (s$^{-1}$) & E$_{f}$ (keV)       & J$_{f}\pi$ & E$_\gamma$ (keV) & Multipolarity & $\lambda_{if}^{\rm exp}$ (s$^{-1}$)     & $\lambda_{if}^{\rm SM}$ (s$^{-1}$)] \\
\hline
0.0     & 0+ & *4.54E-01 & --      & -- & --        & --    & --                 & --            \\
146.36  & 3+ & *2.00E-04 & 0.0     & 0+ & 146.36    & M3    & 1.61E-04          & 1.95E-04     \\
461.00  & 1+ & 3.94E-03 & 0.0     & 0+ & 461.00    & M1    & 1.33E+11          & 6.08E+10     \\
        &    &  & 146.36  & 3+ & 314.64    & E2    & \textless6.67E+08 & 2.07E+07     \\
665.56  & 1+ & 2.07E-02 & 0.0     & 0+ & 665.55    & M1    & 6.33E+10          & 2.44E+10     \\
        &    &  & 146.36  & 3+ & 519.19    & E2    & 5.07E+09          & 6.55E+08     \\
        &    &  & 461.00  & 1+ & 204.55    & E2    & 6.96E+09          & 7.60E+06     \\
1230.26 & 2+ & 5.51E-03 & 0.0     & 0+ & 1230.24   & E2    & 9.69E+08          & 7.47E+08     \\
        &    &  & 146.36  & 3+ & 1083.88   & E2    & 1.65E+10          & 2.96E+09     \\
        &    &  & 461.00  & 1+ & 769.25    & E2    & 1.94E+10          & 5.33E+10     \\
        &    &  & 665.56  & 1+ & 564.68    & E2    & 1.43E+10          & 1.98E+09     \\
1887.14 & 2+ & 1.29E-03 & 0.0     & 0+ & 1887.10   & E2    & \textless2.94E+10 & 5.02E+09     \\
        &    &  & 146.36  & 3+ & 1740.74   & E2    & 2.02E+11          & 1.72E+11     \\
        &    &  & 461.00  & 1+ & 1426.10   & E2    & 3.26E+11          & 3.26E+11     \\
        &    &  & 665.56  & 1+ & 1221.55   & E2    & 9.79E+09          & 6.95E+09     \\
        &    &  & 1230.26 & 2+ & 656.86    & E2    & 9.79E+09          & 2.64E+09     \\
        \hline\hline
\end{tabular}
\end{table*}

\end{document}